\documentclass[aps,twocolumn,showpacs]{revtex4}

\usepackage{graphicx}
\usepackage{epsfig}
\usepackage{amssymb}
\usepackage{amsmath}
\usepackage{psfrag}
\usepackage{color}

\newcommand{\rg}{{\bf r}}
\newcommand{\Eg}{{\bf E}}

\newcommand{\Gg}{{\bf G}}
\newcommand{\pg}{{\bf p}}

\newcommand{\eg}{{\bf e}}

\newcommand{\dd}{{\mathrm{d}}}

\newcommand{\Imag}{\mathrm{Im} \, }

\newcommand{\lcoh}{\ell_{\mathrm{coh}}}

\begin{document}

\title{Spatial coherence in complex photonic and plasmonic systems}
\author{A. Caz\'e, R. Pierrat and R. Carminati}
\email{remi.carminati@espci.fr}
\affiliation{Institut Langevin, ESPCI ParisTech, CNRS, 1 rue Jussieu,
75238 Paris Cedex 05, France}

\begin{abstract}
The concept of cross density of states characterizes the intrinsic spatial coherence of complex
photonic or plasmonic systems, independently on the illumination conditions. Using this tool and the
associated intrinsic coherence length, we demonstrate unambiguously the spatial squeezing of eigenmodes
on disordered fractal metallic films, thus clarifying a basic issue in plasmonics.
\end{abstract}


\maketitle

%Introduction
The optical properties of nanostructured materials have attracted a lot of attention,
due to their potential for light concentration and transport at subwavelength 
scales~\cite{KreibigBook,NovotnyBook}. New possibilities have emerged,
for the design of efficient sources and absorbers of visible and near-infrared radiation, or for optical storage and
information processing with ultrahigh spatial density. Metallic nanostructures benefit from the excitation of surface
plasmons that permit concentration at ultra-small length scales and ultra-fast time scales~\cite{Stockman2011}. 
Disordered media also offer the possibility to build up spatially localized modes (e.g. by the process of Anderson localization)~\cite{ShengBook}. 
Light concentration and transport at subwavelength scales encompass a broad range of processes, 
including coherent control at the nanoscale~\cite{Stockman2007},
enhancement of light-matter interaction in weak and strong coupling 
regimes~\cite{Krachmalnicoff2010,Lodahl2010,Sapienza2011,Chang2007,Bellessa2012}, superradiance~\cite{FJ2012},
enhancement of non-radiative energy transfer~\cite{Vahid2007}, or light focusing beyond the diffraction 
limit~\cite{Stockman2008,Sentenac2008,Mosk2010}.
The spatial extent of eigenmodes is of central importance, since it characterizes the ability of the system
to support concentrated or delocalized excitations. It drives, e.g., 
the coherence length of surface plasmons~\cite{Bellessa2012,Schattschneider2005,DeWilde2006,Krenn2011,Kociak2012},
the range of non-radiative energy transfer~\cite{Vincent2011,Castanie2012}, or the lower 
limit for spatial focusing by time reversal or phase conjugation~\cite{DeRosny2002,Pierrat2007,Lerosey2007}. 
The trade-off between localized and delocalized excitations is also a central issue for the understanding and the control of the
optical properties of disordered fractal metallic films~\cite{ShalaevBook}.
In this Letter, we introduce the Cross Density Of States (CDOS) as a quantity that characterizes the overall spatial extent of eigenmodes,
and use it to address the spatial localization of light on disordered fractal metallic films.
We demonstrate unambiguously the spatial squeezing of eigenmodes close to the percolation threshold, thus providing
a theoretical basis to clarify a
controversial issue in plasmonics~\cite{Krachmalnicoff2010,Stockman2001,Shalaev2003,Cao2006}. This also
illustrates the relevance of the CDOS to characterize the intrinsic spatial coherence in photonics and plasmonics systems.

%Introduction of CDOS
In order to characterize the intrinsic spatial coherence of complex photonic or plasmonic system at a given frequency $\omega$, 
we introduce a two-point quantity $\rho(\rg,\rg^\prime, \omega)$ that we will refer to as CDOS,
defined as
\begin{equation}
\rho(\rg,\rg^\prime, \omega) = \frac{2 \omega}{\pi c^2} \Imag \left [ \mathrm{Tr} \, \Gg(\rg,\rg^\prime,\omega) \right ] \ .
\label{eq:CDOS}
\end{equation}
In this expression, $c$ is the speed of light in vacuum, 
$\Gg(\rg,\rg^\prime,\omega)$ is the electric dyadic Green function that connects the electric field at point $\rg$ 
to an electric-dipole source $\pg$ at point $\rg^\prime$ through the relation $\Eg(\rg) = \mu_0 \, \omega^2 \, \Gg(\rg,\rg^\prime,\omega) {\bf p}$,
and $\mathrm{Tr}$ denotes the trace of a tensor.

%ImG and spatial coherence
The choice of this quantity as a measure of the intrinsic spatial coherence results from the observation
that the imaginary part of the Green function at two different points appears in a number of situations where the spatial
coherence of random fields (produced by random sources and/or a disordered medium) needs to be 
characterized~\cite{Joulain2005,Weaver-Larose,ShengBook,Friberg2003}. 
The imaginary part of the Green function also describes the process of focusing by time
reversal in a closed cavity~\cite{DeRosny2002,Pierrat2007}. 
The precise definition of the CDOS in Eq.~(\ref{eq:CDOS}) has been chosen so that it reduces to the Local Density Of States (LDOS)
when $\rg$ and $\rg^\prime$ coincide~\cite{NovotnyBook,note}.

%CDOS in terms of modes
The physical picture behind the CDOS is a counting of optical eigenmodes that connect two different points at a given frequency.
In a network picture, the LDOS measures the number of channels crossing at a given point, whereas the CDOS measures the number of channels 
connecting two points.
In order to give a more rigorous basis to this picture, let us first consider the canonical situation of a non-absorbing system (e.g., a nanostructured
material) placed in a closed cavity. In this case, using an orthonormal discrete basis of eigenmodes with eigenfrequencies $\omega_n$ and 
eigenvectors $\eg_n(\rg)$, the CDOS defined by Eq.~(\ref{eq:CDOS}) can be rewritten as~\cite{supplemental}:
\begin{equation}
\rho(\rg,\rg^\prime,\omega) =  \sum_n \mathrm{Tr} \left [ \eg_n^*(\rg^\prime,\omega) \eg_n(\rg,\omega) \right ] \, \delta(\omega-\omega_n) \ .
\label{eq:CDOS_modes}
\end{equation}
This expression explicitly shows that the CDOS sums up all eigenmodes connecting $\rg$ to $\rg^\prime$ at frequency $\omega$,
weighted by their strength at both points $\rg$ and $\rg^\prime$. In the case of an open and/or absorbing system, as that considered in the present study, 
the rigorous introduction of a basis of eigenmodes is more involved. 
Approaches have been developed in the quasi-static limit~\cite{KociakPRB2012}, or based on statistical properties of the spectral decomposition
of non-Hermitian matrices~\cite{Markel2006}. 
Assuming weak leakage, one can also use a phenomenological approach in which quasi-modes are introduced by broadening the eigenmodes 
using a linewidth $\gamma_n$. This results in an expansion similar to (\ref{eq:CDOS_modes}) with a Lorentzian lineshape replacing the delta 
function~\cite{supplemental}. This generalizes the physical picture to lossy systems. 
Nevertheless, it is important to note that all calculations presented in this Letter are performed using Eq.~(\ref{eq:CDOS}), in which the correct counting 
of modes is implicit, without referring to a basis of eigenmodes.

%Introduction to disordered metal films
We shall now show that the concept of CDOS allows us to clarify an important issue in nanophotonics
concerning light scattering and localization in disordered fractal metallic films. 
These peculiar structures exhibit optical properties that strongly differ from those of bulk metals or ensembles of isolated 
nanoparticles~\cite{ShalaevBook}. In particular, the multiscale geometry of percolation clusters induces long-range correlations that make simple models
(e.g., white-noise potentials or homogenization procedures) invalid. 
The interplay between surface-plasmon resonances and multiple scattering by the 
fractal percolation clusters leads to spatial concentration of light in subwavelength areas 
(hot spots)~\cite{Gresillon1999,Laverdant2008}. 
The theoretical description of this phenomenon has been the subject of a controversy. Using a scaling theory in the quasi-static limit, a mechanism based on Anderson localization has been put forward [38]. Anderson localization on percolating systems for electronic (quantum) transport leads to a clear transition between the localized and de- localized regimes [39]. For light scattering on percolating metallic systems, a theoretical analysis has proved the existence of localized modes characterized by algebraic rather than by exponential spatial confinement, and that can be coupled to radiation [35]. Numerical simulations on planar random composites have even shown that localized and delocalized plasmonic eigenmodes could coexist [26]. This has been confirmed by measurements and computations of intensity fluctuations in the near field [27, 28], that have also indicated that localized modes should dominate around the percolation threshold (but not exactly at percolation). More recently, measurements of near-field LDOS statistics have confirmed the existence of spatially localized modes in the regime dominated by fractal clusters (close to the percolation threshold) [6].
%The theoretical description of this phenomenon
%has been the subject of a controversy. Using a scaling theory in the quasi-static limit, a mechanism based on Anderson 
%localization has been put forward~\cite{Shalaev1999}. Anderson localization on percolating systems for electronic (quantum)
%transport leads to a clear transition between the localized and delocalized regimes~\cite{Chang-Soukoulis}.
%Nevertheless, one might expect for light (or more generally electromagnetic waves) a different picture due to the inherent long-range interactions.
%Indeed, numerical simulations on planar random composites have shown that localized and delocalized plasmonic eigenmodes could 
%coexist~\cite{Stockman2001}. This has been confirmed by measurements and computations of intensity fluctuations in the near 
%field~\cite{Cao2006,Shalaev2003}, that have also indicated that localized modes should dominate around the percolation threshold 
%(but not exactly at percolation). A theoretical analysis including full retardation has proved the existence of localized modes characterized
% by algebraic rather than by exponential spatial confinement, another difference with Anderson localization of electronic quantum 
%waves~\cite{Markel2006}. More recently, measurements of near-field LDOS statistics have confirmed the existence of spatially localized modes
%in the regime dominated by fractal clusters (close to the percolation threshold)~\cite{Krachmalnicoff2010}. 
For nanophotonics, a major issue
is the description of the overall spatial extent of {\it the full set} of eigenmodes whatever the underlying mechanism 
(regarding this issue, the coexistence of localized and delocalized modes is not a central point). Spatial coherence
and the concept of CDOS appear as a natural tool to address this issue, as we shall see. We will introduce the intrinsic coherence 
length as a measure of this overall spatial extent. This gives a new point of view for the description of light localization on disordered metallic films.

\begin{figure}
\begin{center}
\includegraphics[width=8cm]{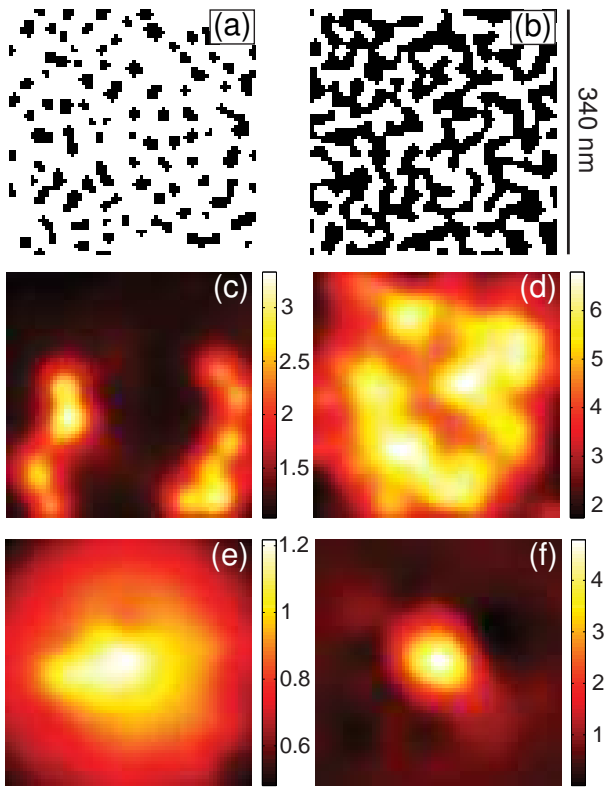}
\caption{\label{fig:maps} (Color online) (a),(b): Geometry of the disordered films generated numerically (with gold in black color).
(a): $f=20\%$, (b): $f=50\%$. (c),(d): Maps of the normalized LDOS $\rho(\rg,\omega)/\rho_0(\omega)$
calculated in a plane at a distance $z=40$ nm above the film surface, $\rho_0(\omega)$ being the LDOS in vacuum .
(e),(f): Maps of the normalized CDOS $\rho(\rg,\rg^\prime,\omega)/\rho_0(\omega)$ with $\rg^\prime$ fixed at the center of the sample.
$\lambda = 780$ nm.  }
\end{center}
\vspace{-0.8cm}
\end{figure}

%Results for disordered metal films
The CDOS can be calculated numerically using exact three-dimensional simulations. We summarize the procedure 
that is fully described in Ref.~\cite{Caze2012}.
Semi-continuous gold films are generated using a Kinetic Monte-Carlo algorithm, reproducing the geometrical features of real films.
Typical realizations of films are shown in the top row in Fig.~\ref{fig:maps} (with gold in black color), each film being 5 nm thick and lying in free space.
To calculate the electric dyadic Green function $\Gg(\rg,\rg^\prime,\omega)$, we need to calculate the electric field at a position $\rg$ 
generated by a point electric-dipole source $\pg$ located at position $\rg^\prime$. To proceed, we solve numerically
the Lippmann-Schwinger equation:
\begin{equation}
\Eg(\rg) = \Eg_0(\rg) + \frac{\omega^2}{c^2}  \int [\epsilon(\rg^\prime,\omega) - 1]\Gg_0(\rg,\rg^\prime,\omega) \Eg(\rg^\prime) \mathrm{d}^3 r^\prime
\end{equation}
where $\Gg_0$ is the vacuum dyadic Green function, $\Eg_0(\rg)$ the incident field and
$\epsilon(\rg,\omega)$ the dielectric function.
The full dyadic Green function is deduced from $\Eg(\rg)=\mu_0 \omega^2 \Gg(\rg,\rg^\prime,\omega) \pg$.
The computation of the LDOS and CDOS follows from Eq.~(\ref{eq:CDOS}).

We show in Fig.~\ref{fig:maps} the LDOS maps (middle row) and CDOS maps (bottom row) computed in a plane
at a distance $z=40$ nm above two different films (shown in the top row) corresponding to two different 
regimes. For $f=20\%$ (left column), the film is composed of isolated nanoparticles 
whereas for $f=50\%$ (right column) the film is slightly below the percolation threshold 
(from numerical simulations, this threshold is estimated at $f=53\%$), a regime in which fractal clusters 
dominate (multiscale resonant regime)~\cite{ShalaevBook,Krachmalnicoff2010,Caze2012}. 
Before studying spatial coherence and the extent of eigenmodes based on the CDOS, 
let us summarize here the main features of the LDOS maps~\cite{Castanie2012,Caze2012}.
For low surface fraction (left column), LDOS peaks are observed on top of isolated nanoparticles that are resonant at the
observation wavelength. A correspondence between LDOS peaks and the position of one or several nanoparticles
is easily made. For a different observation wavelength (not shown for brevity), particles can switch on or off resonance 
and the position of the LDOS peaks changes, but remain attached to individual particles. The sample behaves as a collection
of individual nanoparticles with well identified surface plasmon resonances.
In the multiscale resonant regime (right column), the LDOS structure is more complex. There is no obvious correspondence between the film topography 
(composed of fractal clusters in which the concept of individual nanoparticles becomes meaningless) and the localized field enhancements
responsible for LDOS fluctuations. This is a known feature of fractal metallic films~\cite{Stockman2001,Gresillon1999,Laverdant2008,Greffet2012}.

The maps of the CDOS $\rho(\rg,\rg^\prime,\omega)$ (bottom row in Fig.~\ref{fig:maps}) are displayed versus $\rg$ for a fixed 
position $\rg^\prime$ (chosen at the center of the sample).
Their meaning can be understood as follows: They display the ability of a point $\rg$ at a given distance from the center point
$\rg^\prime$ to be connected to this center point by the underlying structure of the optical eigenmodes. For example, a large CDOS 
(larger than the vacuum CDOS) would allow two quantum emitters at $\rg$ and $\rg^\prime$ to couple efficiently.
It would also ensure coherent (correlated) fluctuations of the light fields at $\rg$ and $\rg^\prime$ under thermal excitation~\cite{Joulain2005}. 
The CDOS also allows one to discriminate between two hot spots at $\rg$ and $\rg^\prime$ that belong to the same eigenmode 
(or that are connected by at least one eigenmode), or that are completely independent. 
Last but not least, since the CDOS implicitly sums up the spatial extent of the full set of eigenmodes,
it appears as a natural tool to describe the overall spatial localization in the multiscale resonant regime.
It is striking to see that the extent of the CDOS in the multiscale resonant regime (Fig.~\ref{fig:maps}f) is reduced to a smaller range
compared to the case of a film composed of isolated nanoparticles (Fig.~\ref{fig:maps}e). 
The reduction of the extent of the CDOS clearly demonstrates an overall spatial squeezing 
of the eigenmodes close to the percolation threshold (remember that the CDOS is implicitly a weighted sum over the full set of eigenmodes). 
Let us stress that the approach based on the CDOS gives a non-ambiguous description of this overall spatial squeezing,
whatever the underlying mechanism. It is based on a concept
implicitly related to field-field spatial correlations as in classical spatial coherence theory, that seems to carry sufficient
information to describe one of the most striking features in the optics of disordered fractal metallic films.

\begin{figure}
\begin{center}
\includegraphics[width=8cm]{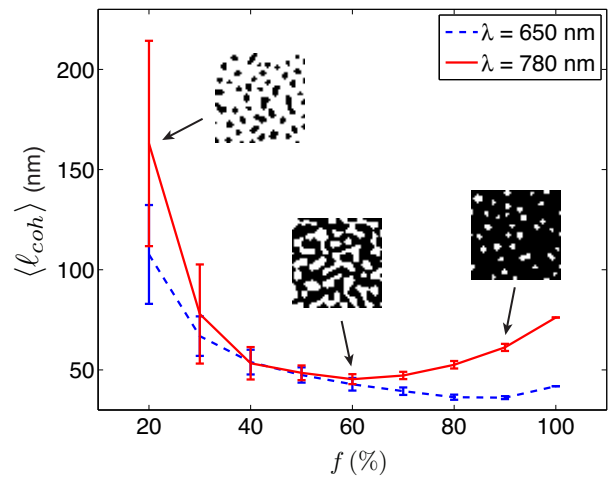}
\caption{\label{fig:mdos_f} (Color online) Averaged value (solid line) and variance
(error bars) of the intrinsic coherence length $\lcoh$ calculated at a distance $z=40$ nm above a disordered film, versus the gold surface fraction $f$. Inset: Typical film geometries (black color corresponds to gold).}
\end{center}
\vspace{-0.7cm}
\end{figure}

In order to quantify the overall reduction of the spatial extent of eigenmodes in the multiscale resonant regime, we
introduce an intrinsic coherence length $\lcoh$, defined from the width of the CDOS.
More precisely, fixing $\rg^\prime$ at the center of the sample, we use polar coordinates in the plane $z=40$ nm
parallel to the sample mean surface to write
$\rho(\rg,\rg^\prime,\omega) = \rho(R,\theta,\omega)$ with $R=|\rg-\rg^\prime]$ and define an angularly-averaged CDOS 
$\bar{\rho}(R,\omega)=(2\pi)^{-1}\int_{0}^{2\pi} \rho(R,\theta,\omega) \dd \theta$. The intrinsic coherence length $\lcoh$ is defined
as the half width at half maximum of $\bar{\rho}(R,\omega)$ considered as a function of $R$.
It is important to note that $\lcoh$ is not necessarily the size of the hot spots observed on the surface, since a given
eigenmode can be composed of several hot spots. Two different hot spots separated by a distance smaller than $\lcoh$ can
be intrinsically connected (meaning that they are connected by at least one eigenmode). The ability
to clarify this distinction between eigenmodes and hot spots is en essential feature of the CDOS.
The averaged value of $\langle \lcoh \rangle$ (solid line) and its variance $\mathrm{Var}(\lcoh)$ (error bars) are shown in 
Fig.~\ref{fig:mdos_f} versus the film surface fraction for two wavelengths $\lambda=650$ nm and $\lambda=780$ nm. 
Both quantities are calculated using a statistical ensemble of realizations of disordered films generated numerically (the error bars
indicate the real variance of $\lcoh$, and not computations errors due to lack of numerical convergence, the latter being ensured 
by a sufficiently large set of realizations). For both wavelengths, the average value $\langle\lcoh\rangle$ is significantly smaller near the percolation 
threshold than for lower filling fractions. This unambiguously demonstrates the overall spatial squeezing of eigenmodes in the regime dominated by fractal clusters,
with a stronger squeezing at $\lambda=780$ nm where more pronounced resonances occur~\cite{ShalaevBook}.
The curve for $\lambda=780$ nm even shows a minimum near the percolation threshold.
Our approach provides a theoretical description of the experiment in Ref.~\cite{Krachmalnicoff2010}, although in 
this study, the inverse participation ratio was used to
connect qualitatively the spatial extent of eigenmodes to the variance of the LDOS fluctuations. Therefore only a qualitative comparison 
with the curve in Fig.~\ref{fig:mdos_f} is possible (the inverse participation ratio and the intrinsic coherence length cannot be compared directly).
Moreover, the precise shape of the calculated curves might also be influenced by finite-size effects inherent to the numerical simulation. 
The behavior of $\mathrm{Var}(\lcoh)$ is also instructive. Strong fluctuations are observed in the regime of isolated nanoparticles. In this
regime, optical modes attached to a single particle and delocalized modes are observed, which is a difference with the known behavior
in quantum electronic transport~\cite{Chang-Soukoulis}. The strong fluctuations
reflect the fluctuations in the interparticle distance. Conversely, in the multiscale resonant regime, 
the reduction of the fluctuations reinforces the
assumption of a mechanism based on collective interactions that involve the sample as a whole.

%CONCLUSION
In summary, we have shown that the CDOS characterizes the intrinsic spatial coherence of a
photonic or plasmonic system, independently on the illumination conditions.  Using this concept, 
we have demonstrated unambiguously the spatial squeezing of plasmonic eigenmodes
on disordered fractal metallic films close to the percolation threshold. This clarifies
a basic issue in plasmonics concerning the description of the optical properties of these films. 
This illustrates the relevance of the CDOS in the study of spatial coherence in photonics and plasmonics systems, and
more generally in wave physics.

We acknowledge Y.~De Wilde, M.~Kociak, V.~Krachmalnicoff and R. Sapienza for stimulating discussions.

%%%%%%%%%%%%%%

\end{document}